\documentclass[9pt, conference]{IEEEtran}
\IEEEoverridecommandlockouts
\usepackage{cite}
\usepackage{amsmath,amssymb,amsfonts}
\usepackage{algorithmic}
\usepackage{graphicx}
\usepackage{textcomp}
\usepackage{xcolor}
\usepackage{cite,url}
\usepackage{color}
\usepackage{multirow, makecell}
\usepackage{kotex}
\usepackage{comment}
\usepackage{arydshln, wasysym, mathtools}
\usepackage{hyperref}
\hypersetup{
    colorlinks=true,
    linkcolor=blue,
    filecolor=magenta,      
    urlcolor=cyan,
}
\def\BibTeX{{\rm B\kern-.05em{\sc i\kern-.025em b}\kern-.08em
    T\kern-.1667em\lower.7ex\hbox{E}\kern-.125emX}}
\begin{document}

\title{D3RM: A Discrete Denoising Diffusion Refinement Model for Piano Transcription 
\thanks{This work was supported by the National Research Foundation of Korea (NRF) grant funded by the Korea government (MSIT) (No. 
RS-2023-NR077289,  No. RS-2024-00358448) and Development of Broadcasting Monitoring Technology for the Settlement and Distribution of Literary Arts (No. RS-2023-00270043)}
}

\author{\IEEEauthorblockN{Hounsu Kim}
\IEEEauthorblockA{\textit{Graduate School of Culture Technology} \\
\textit{KAIST}\\
Daejeon, South Korea \\
hanshounsu@kaist.ac.kr}
\and
\IEEEauthorblockN{Taegyun Kwon}
\IEEEauthorblockA{\textit{Graduate School of Culture Technology} \\
\textit{KAIST}\\
Daejeon, South Korea \\
ilcobo2@kaist.ac.kr}
\and
\IEEEauthorblockN{Juhan Nam}
\IEEEauthorblockA{\textit{Graduate School of Culture Technology} \\
\textit{KAIST}\\
Daejeon, South Korea \\
juhan.nam@kaist.ac.kr}

% \and
% \and
% \and
}

\maketitle

\begin{abstract}
Diffusion models have been widely used in the generative domain due to their convincing performance in modeling complex data distributions. Moreover, they have shown competitive results on discriminative tasks, such as image segmentation. While diffusion models have also been explored for automatic music transcription, their performance has yet to reach a competitive level. In this paper, we focus on discrete diffusion model's refinement capabilities and present a novel architecture for piano transcription. Our model utilizes Neighborhood Attention layers as the denoising module, gradually predicting the target high-resolution piano roll, conditioned on the finetuned features of a pretrained acoustic model. To further enhance refinement, we devise a novel strategy which applies distinct transition states during training and inference stage of discrete diffusion models. Experiments on the MAESTRO dataset show that our approach outperforms previous diffusion-based piano transcription models and the baseline model in terms of F1 score. Our code is available on the accompanying website \footnote{\href{https://github.com/hanshounsu/d3rm}{https://github.com/hanshounsu/d3rm}}.
\end{abstract}


\section{Introduction}
Piano transcription involves converting the acoustic signals of piano performances into individual notes. These notes are usually represented in a MIDI format, containing pitch information and the start and end timings (\textit{onset} and \textit{offset}) of each note. While recent approaches have achieved near-perfect performance \cite{hft2023, yan2024scoring}, most previous work have relied on feed-forward models. This means the \textit{final} prediction for each note is made independently without considering the \textit{final} predictions of other notes. In other words, the model does not effectively consider other neighboring notes when predicting each individual note.

Recent advancements in deep generative models mainly utilize diffusion models, which progressively denoise data initially sampled from a stationary distribution. Continuous domain diffusion is used for continuous data, such as in image and audio generation tasks where it operates in the pixel and sample domain respectively. Discrete domain diffusion, on the other hand, handles discrete data, often in the form of tokens. Discrete diffusion models have recently proved their effectiveness in generative tasks utilizing VQ-VAEs \cite{ju2024naturalspeech}. In addition to their contributions to generative tasks, diffusion models have shown competitive performance in discriminative tasks such as image segmentation\cite{chen2023analog, wang2023dformer}. Building on these foundings, we explore the potential of utilizing discrete diffusion models for piano transcription tasks for the first time. We hypothesize that the discrete denoising module not only predicts the states of each pixel in the piano roll gradually, but also \textit{refines} the previously predicted states, which indicates that the \textit{final} prediction of each note take into account the \textit{final} predictions of other notes more effectively than feedforward approaches.

Previous research explored the use of continuous domain diffusion for transcription \cite{cheuk2023diffroll}, but it failed to achieve performance comparable to current state-of-the-art (SOTA) methods. We address the shortcomings by utilizing a discrete diffusion model, which includes a conditioning encoder, based on a pretrained acoustic model, and a denoising decoder composed primarily of Neighborhood Attention (NA) layers \cite{hassani2023neighborhood}. \def\hide{Additionally, considering the local cohesiveness between the features of the acoustic model and the denoising module, we apply cross-attending NA layers to condition the acoustic features for the denoising decoder.} Additionally, we introduce a novel strategy involving separate transition states for the training and inference stages of discrete diffusion models. Comparing to the baseline, which is the acoustic model trained with a Focal loss, we demonstrate that utilizing the diffusion-based approach is effective for piano transcription.

\section{Related works}
\label{sec:related works}

\subsection{Previous Piano Transcription Models}
Most piano transcription models have utilized feed-forward deep neural networks to estimate notes from audio. Typically, the automatic music transcription (AMT) frameworks are optimized using either frame-wise loss \cite{hawthorne2017onsets, kong2021high, Wei2022HPPNet, kwon2024PAR, hft2023} or a seq2seq framework \cite{hawthorne2021seq2seq}. In addition, \cite{yan2024scoring} recently achieved state-of-the-art results using a Semi-CRF framework. In terms of model architecture, CNN-RNN-based models have been the most common choice for many years \cite{hawthorne2017onsets, kong2021high, Wei2022HPPNet, wu2024piano, kwon2024PAR}, while transformer-based models \cite{hawthorne2021seq2seq, hft2023, yan2024scoring} have also been explored more recently. Another important factor is the representation of the output. Typically, onsets, frames, and offsets are represented as separate binary arrays \cite{hawthorne2017onsets, kong2021high, Wei2022HPPNet, wu2024piano}. In contrast, we employ a multi-label note states representation, where the state of each frame and pitch is expressed with a single discrete multi-label array. We followed the multi-state label approach used by \cite{kwon2020polyphonic}, where the states are \textit{onset}, \textit{sustain}, \textit{re-onset}, \textit{offset}, and \textit{off}.

\begin{figure}[ht!]
\begin{minipage}[b]{1.0\linewidth}
  \centering
  \centerline{\includegraphics[width=8cm]{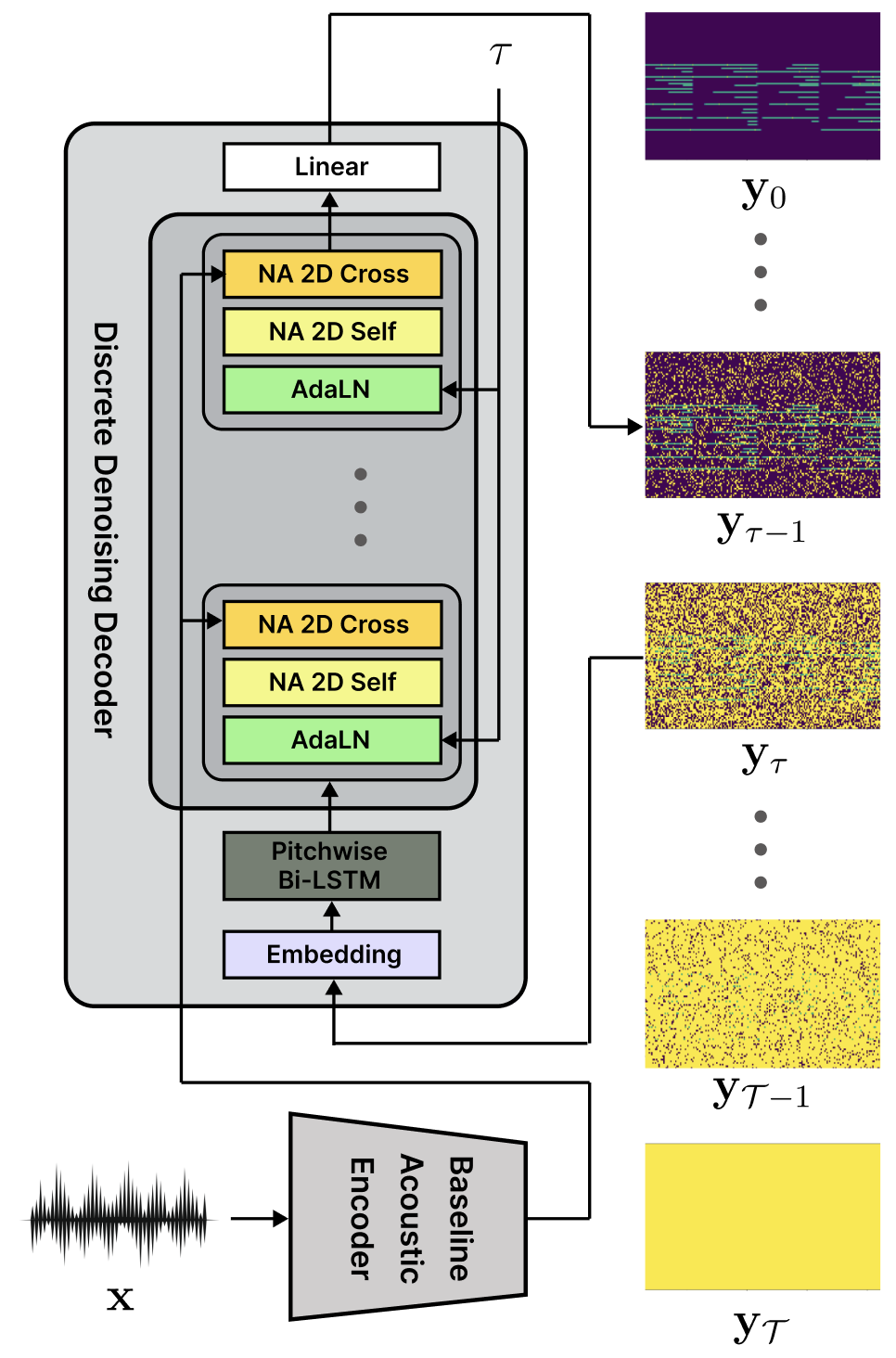}}
 \vspace{-0.3cm}
  % \centerline{(a) Result 1}\medskip
\end{minipage}
\caption{Overall model architecture. \def\hide{$\mathbf{y}_{\tau}$ illustrates the piano roll during the denoising process, and} The yellow pixels within the piano rolls $\mathbf{y}_{\tau}$ represent mask states.} 
\label{fig:model_architecture}
\end{figure}

\subsection{Previous Diffusion-based Approach}
To our knowledge, DiffRoll is currently the only approach that directly utilizes diffusion models for automatic music transcription \cite{cheuk2023diffroll}. DiffRoll is a continuous diffusion model based on the DiffWave\cite{kong2020diffwave} model, where the WaveNet \cite{oord2016wavenet} architecture is employed as the denoising decoder, with diffusion timestep and mel spectrogram conditioned through fully-connected layers and 1D convolutional layers respectively. The piano roll is represented with binary states, \textit{on} and \textit{off}, which are casted to the continuous domain to be input to the diffusion model. \def\hide{The input representation method of DiffRoll is limited due to the lack of states, therefore we incorporated multi-label note states such as \textit{onset}, \textit{sustain}, \textit{re-onset}, \textit{offset}, and \textit{off}, following \cite{kwon2020polyphonic}. Furthermore,} 
One limitation of their approach is the discrepancy between the discrete nature of the note space and the continuous representation. 
In our work, we address this by representing note states as a multi-label discrete variable and applying a discrete diffusion process. 
Furthermore, the audio feature encoder used in the DiffRoll was quite simple, consisting of only a 1D convolutional layer. We reinforced the acoustic encoder of the model by adopting a more structured, existing acoustic encoder.

\subsection{Discrete Diffusion Models}
The diffusion model in the discrete domain was first introduced by \cite{sohl2015binarydiffusion} in the form of binomial diffusion, and has been further expanded to multinomial diffusion models by \cite{hoogeboom2021multinomialdiffusion, austin2021structured}. A transition matrix characterized by the uniform probability of random replacement of the current state with another state, alongside a probability for conversion to a specific [MASK] state, has proven to be particularly effective, as demonstrated in \cite{gu2022vqdiffusion, yang2023diffsound}. As mentioned in \cite{yang2023diffsound}, this choice of transition matrix can be regarded as refining the predicted results during the diffusion process. This insight motivated us to adopt this approach for piano transcription tasks, where the note states undergo refinement.

\subsection{Neighborhood Attention}
Neighborhood Attention (NA)\cite{hassani2023neighborhood} is a sliding window attention mechanism first introduced in the image domain, where each pixel of the image attends to its k-nearest neighbor pixels. By hierachically stacking NA blocks, the attention mechanism can be efficiently applied to high-resolution images while providing locality as an inductive bias. In our work, we employed 2D Neighborhood Attention in the denoising module of the discrete diffusion model, efficiently managing the high resolution of piano rolls. The inductive bias allows each pixel of the piano roll to consider neighboring notes during the denoising process. This can be seen as iterative refinement when the model is trained to allow the state of each pixels to transit to another state during a single denoising step. For parallel computation of Neighborhood Attention, we utilized the \textit{NATTEN} python package\footnote{available at \url{https://github.com/SHI-Labs/NATTEN}}.

\section{Method}
\label{sec:method}

\subsection{Discrete Denoising Decoder}
The denoising decoder takes the roll of a discrete diffusion model, gradually predicting the multi-states of the piano roll and refining the predicted tokens during each diffusion timestep. It is composed of a pitchwise bidirectional LSTM layer followed by stack of NA 2D self-attention blocks, with AdaLN layers in between to condition the diffusion timesteps. The pitchwise bidirectional LSTM operates the LSTM separately for each of the 88 pitches, focusing on the temporal transitions within a single pitch. This layer was demonstrated to be effective in \cite{Wei2022HPPNet, kwon2024PAR}.

\subsection{Acoustic Encoder}
The acoustic encoder receives the audio waveform and provides appropriate conditioning features to the denoising decoder. We adapted the HPPNet \cite{Wei2022HPPNet} with a few modifications as our baseline. The HPPNet takes a Constant-Q Transform (CQT) as input, which is processed by a stack of 2D convolution and harmonic-dilated convolution (HDC) layers. The output of this convolution stack is then passed through FG-LSTM, which is a pitchwise LSTM layer. In our modified model, we used a multi-scale mel-like spectrogram, where each frequency bin is mapped to logarithmically spaced bins aligned with MIDI frequencies. We also increased the number of HDC layers from 1 to 3, employed a single CNN-HDC stack, and adapted a multi-state representation \cite{kwon2020polyphonic}. This model, trained with the Focal loss and without the diffusion process, served as our baseline. \def\hide{Additionally, we used the convolution stack part of this pretrained model as the acoustic encoder for our diffusion model.}

The penultimate output feature of this model has the same height and width dimensions as the piano roll and is sought to contain key features that has local coherence with the expected piano roll output. Therefore, for our diffusion model, we cross-attend these penultimate features of the pretrained baseline model with the denoising decoder using NA, which we call \textit{NA 2d cross}. The overall model architecture is illustrated in Figure \ref{fig:model_architecture}. All encoder parameters are finetuned during the training.

%We leverage a pretrained acoustic model for the encoder, which is originally trained for the same piano transcription task but with focal loss \cite{kwon2024PAR}. \textcolor{red}{This model .... initially extracts CQT from input audio, and expands its hidden dimension through simple conv2d block.} We hereafter refer to this model as \textbf{NAR-HC}, and establish it as the baseline for our approach.

\subsection{Discrete Diffusion for Piano Transcription}
The target label is represented as $\mathbf{y}\in \mathbb{R}^{T\times88\times6}$, where the last dimension is a one-hot vector containing $K=5$ multi-label note states plus a [MASK] state. During the forward process, the label is corrupted by a transition matrix $[\mathbf{Q}_\tau]_{ij}=q(y_\tau^{tp}=j|y_{\tau-1}^{tp}=i)\in\mathbb{R}^{\;6\times6}$ structured as follows:
\begin{equation}\label{eq:1}
    \mathbf{Q}_\tau =
    \begin{bmatrix}
    \alpha_\tau+\beta_\tau & \beta_\tau & \beta_\tau & \dots & 0\\
    \beta_\tau & \alpha_\tau+\beta_\tau & \beta_\tau & \dots & 0\\
    \beta_\tau & \beta_\tau & \alpha_\tau+\beta_\tau & \dots & 0\\
    \vdots & \vdots & \vdots & \ddots & \vdots\\
    \gamma_\tau & \gamma_\tau & \gamma_\tau & \dots & 1\\
    \end{bmatrix}
\end{equation}
where $y_{\tau}^{tp}$ denotes the state of a label $\mathbf{y}$ at position $t\in \{0, 1, ..., T-1\}$, pitch $p \in \{0, 1, ..., 87\}$ and diffusion timestep $\tau \in \{0, 1, ..., \mathcal{T}-1\}$.
The cumulative products $\overline{\mathbf{Q}}_\tau=\mathbf{Q}_\tau \dots \mathbf{Q}_1 $ can then be expressed in a closed form as follows:
\begin{equation}
    \overline{\textbf{Q}}_\tau \mathbf{y}_0[t,p,:] = \overline{\alpha}_\tau \mathbf{y}_0[t,p,:] + (\overline{\gamma}_\tau-\overline{\beta}_\tau)y_{mask} + \overline{\beta}_\tau
\end{equation}
where $y_{mask}$ is the one-hot vector indicating the mask token, $\overline{\alpha}_\tau=\prod_{i=1}^\tau\alpha_i$, $\overline{\gamma}_\tau=1-\prod_{i=1}^\tau(1-\gamma_i)$, and $\overline{\beta}_\tau=(1-\overline{\alpha_\tau}-\overline{\gamma}_\tau)/K$.

The training objective minimizes the KL divergence between the forward process posterior $q(\mathbf{y}_{\tau-1}|\mathbf{y}_\tau, \mathbf{y}_0)$ and the model prediction of the reverse probability distribution $p_\theta(\mathbf{y}_{\tau-1}|\mathbf{y}_\tau, \mathbf{x})$, where $\mathbf{x}$ is the audio waveform. In \cite{austin2021structured, gu2022vqdiffusion} $p_\theta(\mathbf{y}_{\tau-1}|\mathbf{y}_\tau, \mathbf{x})$ is reparametrized as follows:
\begin{equation}\label{eq:3}
    p_\theta(\mathbf{y}_{\tau-1}|\mathbf{y}_\tau, \mathbf{x})= \sum_{\hat{\mathbf{y}}_0 }q(\mathbf{y}_{\tau-1}|\mathbf{y}_\tau, \hat{\mathbf{y}}_0)p_\theta(\hat{\mathbf{y}}_0|\mathbf{y}_\tau, \mathbf{x})
\end{equation}
where $\hat{\mathbf{y}}_0$ is the model's prediction of the target denoised sample. This reparameterization leads to an additional auxiliary denoising objective $\mathcal{L}_{\mathbf{y}_0}= -\text{log}p_\theta(\hat{\mathbf{y}}_0|\mathbf{y}_\tau,\mathbf{x})$ and the corresponding loss weight $\lambda$. The final loss can be described in equation as follows:
\begin{equation}
\begin{split}
    \mathcal{L}_{vlb} &= D_{KL}\big[q(\mathbf{y}_\mathcal{T}|\mathbf{y}_0)||p(\mathbf{y}_\mathcal{T})\big] \\& + \sum^{\mathcal{T}}_{\tau=1}\big[D_{KL}[q(\mathbf{y}_{\tau-1}|\mathbf{y}_{\tau},\mathbf{y}_0)||p_{\theta}(\mathbf{y}_{\tau-1}|\mathbf{y}_{\tau},\mathbf{x})]\big]
\end{split}
\end{equation}
\begin{equation}
    \mathcal{L} = \lambda\mathcal{L}_{\mathbf{y}_0} + \mathcal{L}_{vlb}
\end{equation}
where $p(\mathbf{y}_\mathcal{T})$ is the prior distribution on diffusion timestep $\mathcal{T}$. Note that the posterior terms $q(\mathbf{y}_{\tau-1}|\mathbf{y}_\tau, \mathbf{y}_0)$, $q(\mathbf{y}_{\tau-1}|\mathbf{y}_\tau, \hat{\mathbf{y}}_0)$ are tractable following equation (5) from \cite{gu2022vqdiffusion}. 
\begin{figure}[!t]
\begin{minipage}[b]{1.0\linewidth}
  \centering
  \centerline{\includegraphics[width=8.5cm]{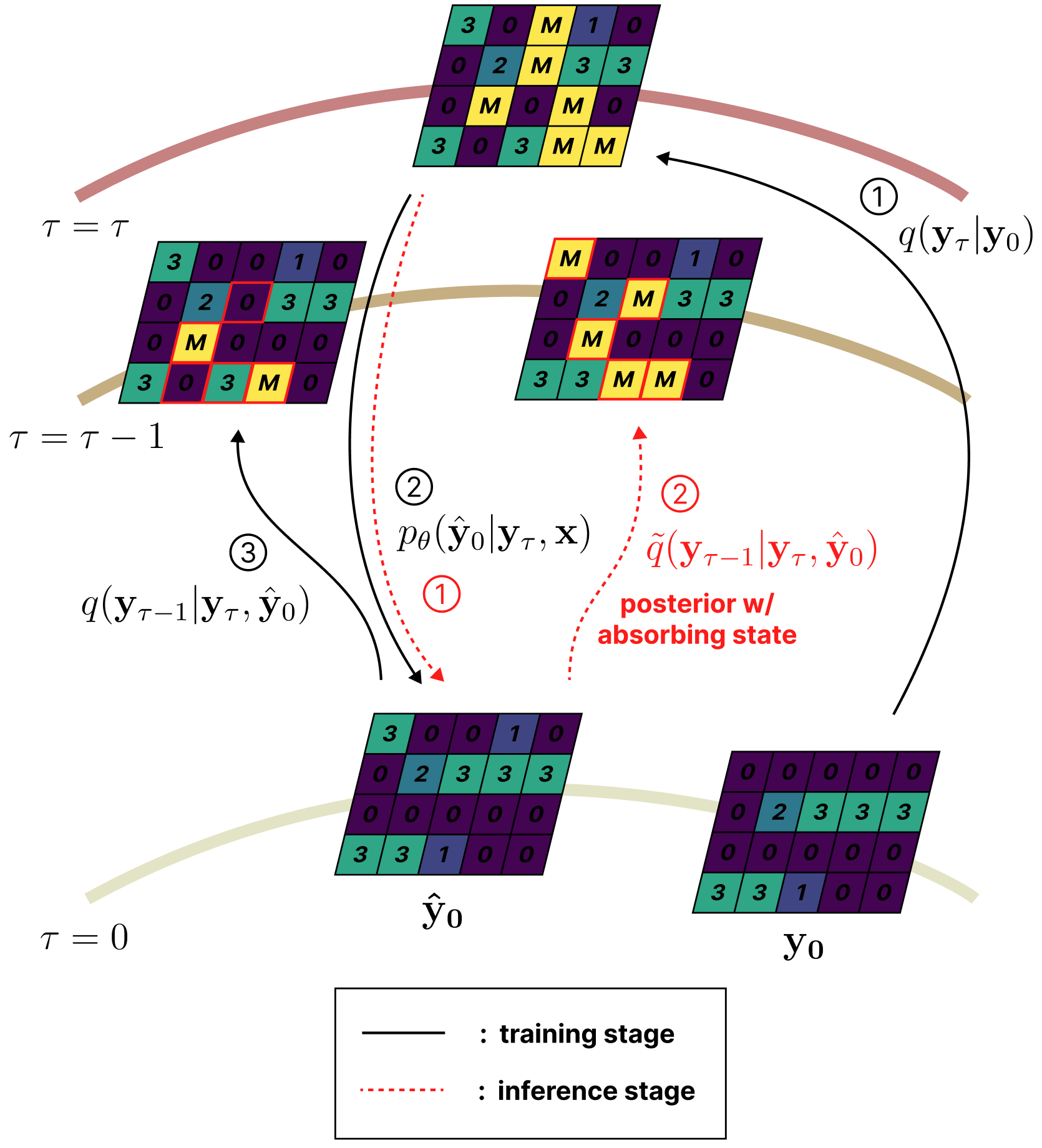}}
%  \vspace{2.0cm}
  % \centerline{(a) Result 1}\medskip
\end{minipage}
\caption{Illustration of the overall discrete diffusion process including the absorbing state for the posterior $\tilde{q}$ in the inference stage. 5 states are visible in this example: off(0), offset(1), onset(2), sustain(3) and mask(M). Details for the posterior $\tilde{q}$ are described in \ref{subsec:absorbing state}.}
\label{fig:diffusion process}
\end{figure}

\subsection{Absorbing State Sampling for Effective Refinement}
\label{subsec:absorbing state}
As shown in (\ref{eq:3}) and illustrated in Figure \ref{fig:diffusion process}, to obtain $p_\theta(\mathbf{y}_{\tau-1}|\mathbf{y}_\tau, \mathbf{x})$, Our model first predicts the clean sample $\hat{\mathbf{y}}_0$, and then applies the transition $q(\mathbf{y}_{\tau-1}|\mathbf{y}_\tau, \hat{\mathbf{y}}_0)$. This method is optimal for the training stage, as the model learns to refine the tokens with probability $\overline{\beta}_\tau$. However, during the sampling stage, the transition replaces the tokens with another token again, which potentially disrupts the refinement process.

To address this, we first simply assume that the model identifies diffusion manifolds as samples with the same total portion of corrupted tokens. Then, instead of corrupting tokens by both replacing \textit{and} masking them with probabilities $\overline{\beta}_\tau$ and $\overline{\gamma}_\tau$ respectively, we can mask the tokens with a total probability of $\overline{\gamma}_\tau + K\overline{\beta}_\tau$, only during the sampling stage ($\tilde{q}(\mathbf{y}_{\tau-1}|\mathbf{y}_{\tau},\mathbf{y}_0)$ in Figure \ref{fig:diffusion process}). This posterior process is identical to the \textit{absorbing state} mentioned in \cite{austin2021structured} and places the sample on the manifold of the desired diffusion timestep. This simple adjustment to the posterior process during sampling leads to improved results. Henceforth, we refer to our proposed model as \textbf{D3RM}, an abbreviation for \textbf{D}iscrete \textbf{D}enoising \textbf{D}iffusion \textbf{R}efinement \textbf{M}odel.

\section{Experiments}
\label{sec:experiments}

\subsection{Dataset and Evaluation}
We trained and evaluated our models with the MAESTRO (V3) dataset\cite{hawthorne2018enabling}, utilizing the provided \{train/valid/test\} splits. To measure the onset and offset accuracy, we employed standard transcription metrics, including a 50ms onset threshold and either a 50ms or 20\% of note length threshold for the offset. We utilized the \textit{mir\_eval} package\cite{mir_eval} for this purpose.

\subsection{Model Configuration}
 Each piano roll state is represented as an embedding vector of size 4. The window size for each of the 8 NA layers is uniformly set to 3, with dilation factors of [1,2,4,8,1,2,4,8]. The hidden dimension of the NA layers is set to 48. Diffusion timesteps are configured with $T=100$, where for the training stage $\overline{\alpha}_\tau$ decreases linearly from 1 to 0, and $\overline{\gamma}_\tau$ increases linearly from $\overline{\gamma}_0 =$ 0 to varying values of $\overline{\gamma}_\mathcal{T} =$ \{0.4, 0.6, 0.8, 0.9, 1.0\}. Note that when $\overline{\gamma}_\mathcal{T} =$ 1.0, $\overline{\beta}_\tau$ remains 0.0 throughout the diffusion process, meaning no substitution occurs, which is equal to the absorbing state. The model does not learn to refine the previous predicted states in this case. The auxiliary loss weight $\lambda$ is set to 0.0005, following \cite{gu2022vqdiffusion}.
 
The model is optimized using the AdamW optimizer with $\beta$s $= [0.9, 0.96]$. The learning rate begins at 0.00045 after a warmup period of 1000 steps, and is reduced by a factor of 0.8 when the diffusion loss does not improve for 25k steps, following \cite{gu2022vqdiffusion}. The acoustic encoder has total 1.5M trainable parameters, and the denoising decoder has 477K parameters. Note that during the inference stage, to reduce overall sampling time, we save the output feature from the encoder at the first diffusion timestep. This saved feature are then fed into the decoder during the remaining timesteps, requiring only a single initial forward pass through the encoder. This significantly accelerates inference, as the lightweight denoising decoder processes data quickly on its own. Model is trained for 400K steps with batch size of 8. We train the model on a single RTX 4090 GPU for 7 days. Inference for the test set took around 3.5 hours.

\section{Results}

\subsection{Main Results}
As depicted in Table \ref{tab:main_results}, our D3RM outperforms the baseline in terms of note F1 score, both with and without offsets. Compared to DiffRoll, D3RM utilizes much fewer parameters and shows a significant improvement. D3RM also outperforms most previous piano transcription models, except for the recent model \cite{yan2024scoring}. \def\hide{We believe that by incorporating a more robust baseline encoder, we could achieve even better results.}

\subsection{Ablation Studies}
We conducted additional experiments on various settings, as shown in Table \ref{tab:ablation}. The top half of the table examines the effects of using the absorbing state sampling (AS sampling), pretrained encoder versus jointly training the encoder from scratch (Enc. Init.), conditioning encoder features via cross-attention or in-context conditioning. The bottom half investigates the impact of linearly increasing $\overline{\gamma}_\tau$ from 0 to varying values of $\overline{\gamma}_\mathcal{T}$. In-context conditioning concatenates the output feature of the acoustic encoder with the noisy piano roll along the channel-wise direction before feeding it into the denoising decoder.

When using the absorbing state ($\overline{\gamma}_\mathcal{T}=1.0$) for the training stage, it demonstrated poorer performance (F1 97.55 $\rightarrow$ 96.82). This suggests that allowing the model to learn transitions between states ($\overline{\gamma}_\mathcal{T} < 1.0$) enables an effective refinement of the piano roll. Our proposed refinement method (AS sampling) also improved the performance (F1 97.21 $\rightarrow$ 97.55) which shows the effectiveness of selecting absorbing state transition matrix for the sampling stage. For the rest of the settings, including training the encoder parameter from scratch, in-context conditioning, and varying $\overline{\gamma}_\mathcal{T}=\{0.4, 0.6, 0.8\}$, the results showed only subtle and unremarkable differences.

\begin{table}[!t]
\begin{center}
 \caption{Test metrics for MAESTRO dataset (V3). Best F1 scores are highlighted in bold.}
\label{tab:main_results}
{\Huge
\renewcommand{\arraystretch}{1.2}
\resizebox{0.48\textwidth}{!}{
\begin{tabular}{lccccccc}
\Xhline{2\arrayrulewidth}
& & \multicolumn{3}{c}{Note Metrics} & \multicolumn{3}{c}{\makecell{Note F1 \\ w\textbackslash offsets (\%)}} \\ \hline
%{\begin{tabular}[c]{@{}c@{}}Note Metrics with\\ Offsets\end{tabular}} \\ \hline
\multicolumn{1}{c}{Model} & \# Param &  P (\%) & R (\%) & F1 (\%) & P (\%) & R (\%) & F1(\%) \\ \hline
% HPT-T\cite{ou2022exploring} &  & 97.88 & 96.72 & 96.77 & 84.13  & 82.31 & 83.20 \\
% Semi-CRFs\cite{yan2021skipping} & & 98.69 & 93.96 & 96.11 & 90.79 & 86.46 & 88.42 \\
% High-resolution\cite{kong2021high} & 20M & & & & & 95.6 & 98.1 & 96.8 & 83.7 & 85.8 & 84.7 \\
HPPNet-sp\cite{Wei2022HPPNet} & 1.2M & 98.45 & 95.95 & 97.18 & 84.88  & 82.76  & 83.80 \\
HFT-transformer\cite{hft2023} & 5.5M & 99.64 & 95.44 & 97.44 & 92.52 & 88.69 & \underline{90.53}  \\
Yan et al.\cite{yan2024scoring} & 12.9M & 99.53 & 97.16 & \textbf{98.32} & 94.61 & 92.39 & \textbf{93.48} \\
% Harmonic Attention\cite{wu2024piano} & & & & & & 98.57 & 96.14 & 97.33 & 85.29 & 83.21 & 84.22 \\
PAR\cite{kwon2024PAR} & 19.7M & 98.49 & 95.60 & 97.01 & 89.21 & 86.68 & 87.88 \\
DiffRoll (k=9)\cite{cheuk2023diffroll} & 86.8M & & & 78.1 & & & \\ \hline
%\Xhline{2\arrayrulewidth}
Baseline & 1.5M & 99.08 & 94.03 & 96.45 & 89.93 & 85.45 & 87.60 \\ \hdashline
D3RM (Proposed) & 2.0M & 98.72 & 96.48 & \underline{97.57} & 91.48 & 89.44 & 90.44 \\
\Xhline{2\arrayrulewidth}
\end{tabular}
} }
\end{center}
\vspace{-0.3cm}
\end{table}

\begin{table}[!t] 
\begin{center}
\caption{Ablation study on D3RM, focusing on AS Sampling, encoder parameter initialization method (Enc. Init.), encoder conditioning method (End. Cond.), and $\gamma_\tau$ scheduling ($\overline{\gamma}_\mathcal{T}$)}
\label{tab:ablation}
{\huge
\renewcommand{\arraystretch}{1.2}
\resizebox{0.48\textwidth}{!}{
\begin{tabular}{cccccc}
\Xhline{2\arrayrulewidth}
% & & & & \makecell{Note \\ Metrics} & \makecell{Note \\ Metrics \\ w\ Offsets} \\ \hline
%{\begin{tabular}[c]{@{}c@{}}Note Metrics with\\ Offsets\end{tabular}} \\ \hline
 AS Sampling & Enc. Init. & Enc. Cond. & $\overline{\gamma}_\mathcal{T}$ &  Note F1 (\%) & \makecell{Note F1 \\ w\textbackslash offsets (\%)} \\ \hline
% HPT-T\cite{ou2022exploring} &  & 97.88 & 96.72 & 96.77 & 84.13  & 82.31 & 83.20 \\
% Semi-CRFs\cite{yan2021skipping} & & 98.69 & 93.96 & 96.11 & 90.79 & 86.46 & 88.42 \\
% High-resolution\cite{kong2021high} & 20M & & & & & 95.6 & 98.1 & 96.8 & 83.7 & 85.8 & 84.7 \\
% HPPNet-sp\cite{Wei2022HPPNet} & 1.2M & & & & & 98.45 & 95.95 & 97.18 & 84.88  & 82.76  & 83.80 \\
% HFT-transformer\cite{hft2023} & 5.5M & & & & & 99.64 & 95.44 & 97.44 & 92.52 & 88.69 & 90.53  \\
% Yan et al.\cite{yan2024scoring} & 12.9M & & & & & 99.53 & 97.16 & \textbf{98.32} & 94.61 & 92.39 & \textbf{93.48} \\
% % Harmonic Attention\cite{wu2024piano} & & & & & & 98.57 & 96.14 & 97.33 & 85.29 & 83.21 & 84.22 \\
% PAR\cite{kwon2024PAR} & 19.7M & & & & & 98.49 & 95.60 & 97.01 & 89.21 & 86.68 & 87.88 \\
% DiffRoll (k=9)\cite{cheuk2023diffroll} & 86.8M & & & & & & & 78.1 & & & \\ \hline
% %\Xhline{2\arrayrulewidth}
% NAR-HC (Baseline) & 1.5M & & & & & 99.08 & 94.03 & 96.45 & 89.93 & 85.45 & 87.60 \\ \hdashline
 $\ocircle$ & Pretrained & Cross & 0.9 & 97.55 & 90.28 \\\hdashline
  $\times$ & Pretrained & Cross & 0.9 & 97.21 & 89.10 \\
  $\ocircle$ & Scratch & Cross & 0.9 & 97.46 & 89.93 \\ 
  $\ocircle$ & Pretrained & In-context & 0.9 & 97.54 & 90.24 \\ \hline
  $\ocircle$ & Pretrained & Cross & 1.0 &  96.82 & 88.33 \\ 
  $\ocircle$ & Pretrained & Cross & 0.8 &  97.51 & 90.10 \\ 
 $\ocircle$ & Pretrained & Cross & 0.6 & \underline{97.57} & \textbf{90.44} \\ 
  $\ocircle$ & Pretrained & Cross & 0.4 &  \textbf{97.58} & \underline{90.37} \\ 
\Xhline{2\arrayrulewidth}
\end{tabular}
}}
\end{center}
\vspace{-0.3cm}
% \vspace{-0.3cm}
\end{table}

\begin{figure}[ht!]
\begin{minipage}[b]{1.0\linewidth}
  \centering
  \centerline{\includegraphics[width=9.5cm]{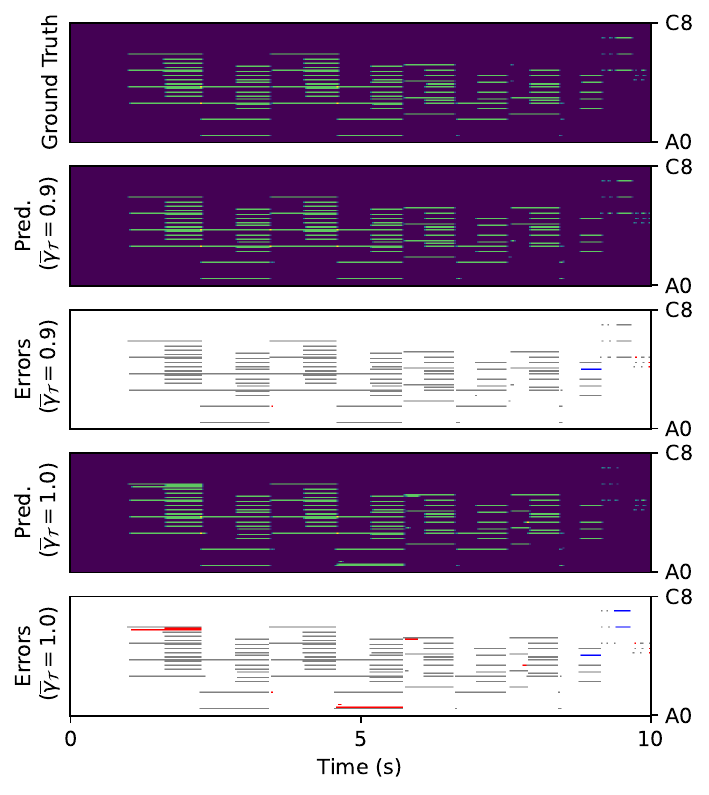}}
 \vspace{-0.3cm}
  % \centerline{(a) Result 1}\medskip
\end{minipage}
\caption[]{Transcription example from a single 10-second sample\footnotemark. In the case where $\overline{\gamma}_\mathcal{T}$ = 0.9, the model is trained with refinement and sampled using AS Sampling. In contrast, for $\overline{\gamma}_\mathcal{T}$ = 1.0, the training and inference stages use the absorbing state, meaning the model does not learn refinement. In the 3rd and 5th row figures, red-colored notes represent mispredicted notes, blue-colored notes indicate unpredicted notes, and grey-colored notes are those correctly predicted. The note errors are determined based on whether each predicted note's onset is correct.}
\label{fig:sample}
\end{figure}

\footnotetext{MIDI-Unprocessed\_12\_R1\_2006\_01-08\_ORIG\_MID--AUDIO\_12\_R1\_2006\_07\_Track07}

\subsection{Empirical Analysis}
To investigate the difference between the models with and without refinement capability ($\overline{\gamma}_\mathcal{T} = 0.9$ vs. $\overline{\gamma}_\mathcal{T} = 1.0$), we visualized the errors of each model and identified the differences. We hypothesized that in the case of the model that has no refinement capability, false note frames predicted during the denoising process may not be corrected. Indeed, we observed errors in the predictions of the model $\overline{\gamma}_\mathcal{T} = 1.0$, which we suspect is a result of this issue. In Figure \ref{fig:sample}, the predictions for $\overline{\gamma}_\mathcal{T} = 1.0$ show several false positive errors, where onsets are located near other note onsets (illustrated with red lines in the last row). These results suggest that the refinement capability is beneficial for correcting false predictions during the diffusion process.

\section{Conclusion}

We demonstrate the potential of discrete denoising diffusion probabilistic models for discriminative tasks by adding a lightweight NA-based denoising module on top of a pre-existing acoustic model.  Additionally, we propose a novel inference strategy which enhances the refinement capability of discrete diffusion models. The results in piano transcription validate the effectiveness of this approach, which could potentially be extended to other domains or even generative tasks. Future directions include the development of a robust objective metric to better evaluate the model's refinement capabilities and the exploration of alternative token representations, such as note-level representations, for discrete token transitions. Finally, we plan to comprehensively assess the generalization performance of this approach across diverse acoustic models through additional experiments.

\section*{Acknowledgment}

During the writing process, ChatGPT 4.0 was minimally utilized solely to ensure grammatical accuracy and thus improve the clarity of the paper.

% \vfill
% \section*{References}
\bibliographystyle{IEEEtran}
\bibliography{refs}

\end{document}